# Suspended graphene membranes to control Au nucleation and growth

Joachim Dahl Thomsen[1], Kate Reidy[1], Thang Pham[1], Julian Klein[1], Anna Osherov[2], Rami Dana[1], Frances M. Ross[1*]

[1]Department of Materials Science and Engineering, Massachusetts Institute of Technology, Cambridge, Massachusetts 02139, USA

[2]Department of Electrical Engineering and Computer Science, Massachusetts Institute of Technology, Cambridge, Massachusetts 02139, USA

*fmross@mit.edu

KEYWORDS: Graphene, 2D materials, nucleation, diffusion, transmission electron microscopy

ABSTRACT

Control of nucleation sites is an important goal in materials growth: nuclei in regular arrays may show emergent photonic or electronic behavior, and once the nuclei coalesce into thin films, the nucleation density influences parameters such as surface roughness, stress, and grain boundary structure. Tailoring substrate properties to control nucleation is therefore a powerful tool for designing functional thin films and nanomaterials. Here, we examine nucleation control for metals deposited on two-dimensional (2D) materials in a situation where substrate effects are absent and heterogeneous nucleation sites are minimized. Through quantification of faceted, epitaxial Au island nucleation on graphene, we show that ultra-low nucleation densities with nuclei several micrometers apart can be achieved on suspended graphene under conditions where we measure the nucleation density to be 2-3 orders of magnitude higher on the adjacent supported substrate. We estimate diffusion distances using nucleation theory and find a strong sensitivity of nucleation and diffusion to suspended graphene thickness. We suggest that nucleation site density control via substrate tuning may act as a platform for applications where non-lithographic patterning of metals on graphene is beneficial.

1. Introduction

In the initial stages of metal deposition, atoms diffuse on the substrate surface and aggregate to form islands.[1,2] In the absence of heterogeneous nucleation sites, these islands form at random locations with a density determined by the atom diffusivity and flux.[1] Deliberate introduction of sites for nucleation is a powerful strategy to introduce plasmonic or electronic functionalities through self-assembly of nanostructure arrays.[3,4,5] Furthermore, as the nuclei coalesce to form a thin film, the nucleation density determines parameters such as film stress, the density and nature of grain boundaries and grain crystallography[6,7,8]. These parameters control the properties of metal thin films in advanced microelectronic, optical, and magnetic devices,[9] making nucleation control critical for design of both thin films and functional nanostructure arrays. The use of 2D materials as the substrate allows for new degrees of control. Large, step-free areas are possible: exfoliated 2D materials can achieve lateral length scales of ~1 mm,[10] while sizes of 2D materials grown by chemical vapor deposition can extend across an entire wafer-scale growth substrate. Furthermore, the weak van der Waals bonding across such extended areas is expected to modify nucleation dramatically. 2D substrates are therefore expected to generate opportunities for controlling nucleation to tailor the structure and properties of both island arrays and coalesced thin films and island arrays for applications such as structural color generation,[4] energy devices,[11] and biosensing.[12]

Graphene is particularly interesting as a nucleation and growth substrate because it has been shown that the morphology and nucleation density of metals on graphene is tunable by controlling the substrate on which the graphene is placed.[2, 13] Metals deposited and then annealed on $SiO_2$-supported mono- or few-layer graphene (both referred to here as Gr) to achieve equilibrium conditions show island number density that depends on the number of graphene layers, with lower density for thicker Gr.[14-17] The mechanism of this effect has been discussed in relation to the observation that surface strain variations and roughness affect Au diffusion on graphite and Gr.[13] Understanding these phenomena in detail would enable nucleation tuning for metals on 2D surfaces and hence optimized strategies for fabricating functional materials.[18, 19]

In order to separate out the effects of strain, roughness, and unintentional nucleation sites, it is useful to consider nucleation of Au on suspended (free-standing) Gr, i.e. Gr that is not in direct contact with a substrate. Suspended Gr (and other suspended 2D materials) offers advantages for understanding metal diffusion and nucleation since eliminating the substrate helps to reveal the intrinsic interaction between metal and 2D material; suspended materials are also promising for applications such as optoelectronics,[20] filtration,[21] DNA sequencing,[22] and as catalysis.[23, 24] However, nucleation phenomena on suspended Gr are obscured by the presence of contamination from sample preparation methods,[25, 26, 27] hydrocarbons and adsorbed water from ambient air,[28] or tape residues used in the exfoliation process[13] which act as heterogeneous nucleation sites. Even on bulk graphite, heterogeneous sites can dominate metal nucleation: The lowest Au nucleation densities on graphite, in the range 1-10 $\mu m^{-2}$, were achieved by cleaving in ultra-high vacuum (UHV) conditions[28, 29] or by annealing air-cleaved graphite in UHV conditions at 600 °C for several hours.[30] Such values match well with reported defect densities in natural graphite.[31-33] Extraneous nucleation sites are particularly critical since Au is extremely mobile on graphene. Density functional theory (DFT) calculations of Au adatom diffusion as a function of graphene layer thickness[34] and Au cluster size (1-4 atoms)[35] have shown that diffusion barriers of Au on Gr are comparable to or less than the thermal energy at room temperature.

Here, we aim to separate the factors that influence nucleation sites. We choose a system, Au deposition on suspended Gr membranes, where substrate effects are eliminated and heterogeneous nucleation sites arising from contamination are minimized. We show that the substrate-free suspended Gr displays Au nucleation density orders of magnitude lower than the nearby, supported regions of the same Gr crystal. We further show that for suspended Gr, the Au nucleation density is thickness-dependent, decreasing with increased layer number and attaining values similar to those seen for Au nucleation on clean bulk graphite.[28-30] These results expand on the previous studies that have examined supported graphene.[13-17] The thickness and substrate dependence suggests a model where surface roughness determines the diffusion characteristics of Au on Gr and hence the nucleation density. We discuss the modulation of nucleation density without lithography of the surface by control of diffusion through substrate engineering.

2. Results and Discussion
2.1. Ultra-low and layer-dependent Au nucleation density on suspended Gr

We first discuss nucleation of Au on Gr and the conditions under which particularly low nucleation density occurs. Figure 1 shows the results of depositing Au on Gr after a sample preparation, cleaning and deposition process described in Methods. The Gr was transferred onto a SiN membrane with holes so that the Gr is free-standing or "suspended" over the holes but "supported" by SiN between holes[18, 19]. An overview of such samples is given in Fig. S1. Figure 1a, b shows transmission electron microscopy (TEM) and scanning TEM (STEM), respectively, of an exfoliated Gr crystal 3 layers (L) in thickness while Fig. 1c shows thicker (10L) Gr. In each image, both suspended and supported regions are visible. The images show a strikingly lower nucleation density ($N$) of Au on suspended Gr compared to supported Gr. In Fig. 1b the SiN-supported region has $N_{3L, supp}=\sim 110$ $\mu m^{-2}$, compared to $N_{3L, susp}=1.6$ $\mu m^{-2}$ in the suspended region. We find reproducibly that $N$ is significantly lower on suspended compared to supported Gr. We also find that thicker Gr has lower $N$: in Fig. 1c a 10L thick suspended Gr area did not nucleate any Au, demonstrating a strong suppression of nucleation density.

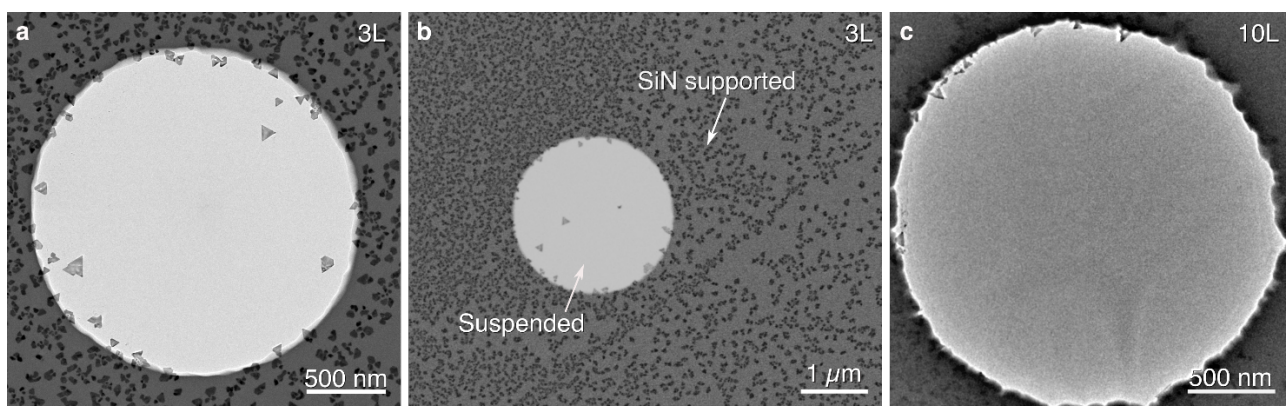

*Figure 1.* Low nucleation density on suspended Gr compared to SiN supported Gr. (a, b) TEM and lower magnification high-angle annular dark field STEM images of 3L Gr with 0.15 nm Au. The contrast in (b) has been inverted to make it easier to compare the images. (c) TEM image of 10L Gr. No nucleation of Au was visible on the suspended areas but rather on the edge of the hole and on the supported regions. Note the weak diffraction contrast in (c) arising from bending of the membrane. The mottled background contrast in (c) is due to contamination of the sample after removal from the UHV system and therefore does not affect the nucleation density measurements.

The low nucleation densities of as-deposited Au shown in Fig. 1, in comparison with higher nucleation densities seen elsewhere,[13-15, 17] emphasize the importance of eliminating contamination arising from the exfoliation, transfer method, or ambient air to avoid distorting measurements of $N$ by extrinsic nucleation sites. Our sample preparation procedure combines UHV deposition conditions, a specific polymer used as handle for the transfer, and sample annealing prior to evaporation, to create suspended Gr samples that are sufficiently clean for these measurements.[19] Several studies have previously shown that areas of clean suspended graphene can also be obtained by methods such as annealing in $H_2/Ar$,[27] laser annealing,[36] and cleaning with activated carbon.[37] We carry out imaging directly after Au deposition (without any post-growth anneal) to capture the nucleation density and island morphology. Note that we do not require (or expect) the samples to remain ultra-clean during imaging, since the contamination is picked up when moving the sample from the UHV system to the TEM[38] and does not modify the Au island parameters.

Having established methods for obtaining ultralow $N$ in suspended Gr, we investigate Au nucleation and diffusion in greater detail using samples in which the transferred Gr crystal is made up of regions of varying thickness. Figure 2 shows measurements as a function of Gr layer number after depositing Au on clean, suspended Gr. The Au has formed well-dispersed, compact, and faceted islands with $N$ clearly varying with layer number. Figure 2a shows images of Au islands nucleated on 1L, 2L, and 4L suspended Gr. Figure 2b shows histograms of the Au island area on the corresponding layer thicknesses across the entire sample. The island shapes vary, but for context, a triangular island with area $6 \cdot 10^3$ nm$^2$ has an edge length of 118 nm. In Fig. 2c, d we plot the layer-dependent average $N$ and the average island area for several samples. We see a decreasing trend for $N$ with $N_{1L}$=7.8 $\mu$m$^{-2}$ as average for monolayer Gr, while for membranes thicker than 5L, $<N_{>5L}$=0.8 $\mu$m$^{-2}$. Measurements of the values of $N$ for thicker membranes are limited by the hole diameter, as in Fig. 1b. However, the values obtained are still comparable to literature values of $N$=1-10 $\mu$m$^{-2}$ on clean bulk graphite with UHV deposition conditions.[28-30] Thicker substrates, as in the very thick graphite in the lower right region of Fig. S2, appear to show values of $N$ that are similar to the value at 5-8L. In the following we, therefore, treat $N_{>5L}=N_{\infty}$=0.8 $\mu$m$^{-2}$ as reference for $N$ on bulk graphite.

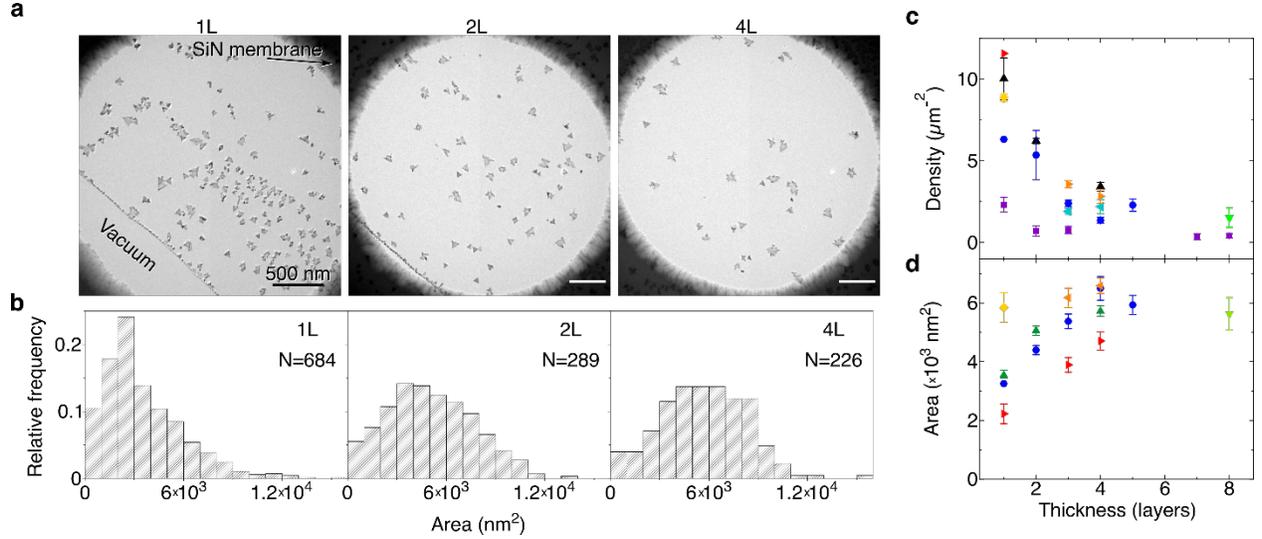

*Figure 2. Layer dependent Au deposition characteristics on Gr. (a) TEM images of Au deposited on suspended 1L, 2L, and 4L Gr that are all from the same sample. (b) Histograms of Au areas from the sample shown in (a). The island areas were measured over several holes, at each of which the Gr thickness is known. The combined areas measured were 72, 39, and 79 µm² for 1L, 2L, and 4L Gr, respectively. (c) Au nucleation density on suspended Gr plotted against Gr thickness. The six different symbol colors/shapes correspond to six different samples. (d) Average island area plotted against Gr thickness for each sample and each thickness. Error bars in (c, d) are the standard errors. Data points without error bars in (c) are measurements from a single suspended area. See Methods for details on nucleation density measurements. One sample in (c) (purple, square data points) was not plotted in (d) due to challenges in measuring the island areas, see Methods. That sample is shown in Figure S2.*

In line with the procedure of previous studies we use mean-field diffusion theory to estimate the diffusion constant of Au on our samples.[13, 14] Since the flux of Au atoms from the evaporator is constant onto all regions of the sample, we estimate the difference in Au diffusivity when comparing results for different layer numbers using[1]

$$N \propto \left(\frac{1}{D}\right)^{\frac{i}{i+2.5}} \quad (1)$$

where $i$ is the critical nucleus size of Au on Gr and $D$ is the Au diffusivity. Using $N_{1L}=7.8$ µm$^{-2}$ and $N_\infty=0.8$ µm$^{-2}$ for the nucleation density on suspended monolayer Gr and bulk graphite, respectively, and assuming $i=1$ and 2 (Ref. 13), results in $\frac{D_{1L}}{D_\infty} = 3 \cdot 10^{-4}$ and $6 \cdot 10^{-3}$, respectively. Hence, we estimate that the Au diffusivity on monolayer suspended Gr is 2-3 orders of magnitude lower than $D_\infty$. We can also compare the Au diffusivity on suspended and SiN-supported Gr. From Fig. 1b we had $N_{3L, supp}=\sim110$ µm$^{-2}$ and $N_{3L}=1.6$ µm$^{-2}$. This results in $\frac{D_{3L,supp}}{D_{3L,susp}} = 4 \cdot 10^{-7}$ and $7 \cdot 10^{-5}$ with $i=1$ and 2, respectively, a difference of 4-6 orders of magnitude between suspended and supported graphene. We can further estimate the characteristic diffusion length $L_D = \frac{1}{\sqrt{N}}$, which gives $L_{D,1L} = 0.34\,\mu m$ and $L_{D,\infty} = 1.12\,\mu m$ for suspended monolayer and graphite, respectively. Such long diffusion lengths for graphite are corroborated by e.g. Fig. 1c, where the absence of nucleated islands indicates a diffusion length larger than the hole radius of ~1 $\mu m$.

In similar vein we can estimate the activation energy for diffusion, $E_d$, by noting that $D \propto \exp\left(-\frac{\Delta E}{kT}\right)$, where k is the Boltzmann constant and $\Delta E$ is the diffusion barrier. Combining this with Eq. 1 we get

$$N \propto \exp\left(\frac{i}{i+2.5}\frac{E_d}{kT}\right). \qquad (2)$$

This allows estimating the change in diffusion barrier as $E_{1L} - E_\infty = \left(\frac{i+2.5}{i}\right)kT\ln\left(\frac{N_{1L}}{N_\infty}\right) =$ 0.21 eV and 0.13 eV for $i$=1 and 2, respectively. The same comparison for supported graphene and SiN suspended graphene results in $E_{3L,\ supp} - E_{3L,\ susp}$ = 0.39 eV and 0.25 eV for $i$=1 and 2, respectively. Absolute values can be estimated by assuming values for $E_\infty$ which has been variously calculated to be 50 meV,[39] or 7-25 meV depending on the diffusion pathway for mono-, bi-, and trilayer graphene.[34] Supplementary Table 1 summarizes the results of our diffusion constant and barrier calculations. It also includes additional data on $SiO_2$ supported graphene that will be presented in the following section.

Thus, the difference in Au diffusivity and diffusion barrier is significantly greater when comparing suspended and supported Gr against suspended monolayer Gr and bulk graphite. Given the strong difference in $N$ between suspended and supported Gr, shown in Fig. 1b, this suggests that areas with different thin film properties can be created through modifying the substrate in certain areas to create suspended regions.

To explain the observed layer dependence of $N$, we first consider a model in which charge transfer between Au and Gr was proposed to be a key factor, based on prior observations of a layer thickness-dependent variation in island size.[15] However, this model applies for islands that are smaller than in our case, since the side lengths of our Au islands are in the range 70-120 nm for 1-8L Gr, significantly larger than the Au islands that typically form on supported Gr.[13-17] This indicates that a mechanism different from charge transfer is needed to explain the thickness-dependent variations in $N$ in our samples. Furthermore, since our $N$ for >5L thick suspended Gr is comparable to reported defect densities in natural graphite,[31-33] this suggests that the factors that increase $N$ for monolayer Gr and thin few-layer Gr crystals are largely eliminated at 5L thickness.

Instead, we suggest that our observed layer dependence on $N$ is consistent with models that are based on a decrease in surface roughness with thickness. Ref. 13 found a larger Au nucleation density on monolayer graphene supported on $SiO_2$ compared to monolayer graphene placed on an intermediary hBN crystal. This was attributed to a decrease in surface roughness and the fact that diffusion is lowered or increased by compressive or tensile strains, respectively.[40, 41] Roughness can increase $N$ because regions with compressive strain, and thus lower Au diffusivity, will act as effective concentrators of adatoms and hence nucleation centers. The root mean square (RMS) surface roughness of monolayer Gr supported on $SiO_2$ or SiN has been reported to be between 0.19-0.35 nm,[42-45] with this relatively high value the result of the substrate not being atomically smooth. Thicker crystals of hBN (~5-20 nm) are flatter with measured RMS surface roughness in the range 0.03-0.25 nm.[45-48] On the other hand, suspended monolayer Gr is known to exhibit even lower intrinsic roughness with RMS values on the order of 0.1 nm,[49-52] and furthermore, the roughness of suspended few-layer Gr decreases as the number of layers increases.[53] Thus, situations of lower roughness correspond to the circumstances of smaller $N$. This is consistent with our observations of apparently increased Au diffusivity both on suspended Gr compared to supported Gr and as the layer thickness increases.

It is also possible, based on theoretical considerations that the thickness dependent Au diffusion properties could have an intrinsic origin since DFT calculations have shown that binding energies for Au on graphene are layer-dependent and converge towards a bulk value for increasing layer thickness.[34] Whether roughness-dependent diffusivity or thickness-dependent binding energies are the origin of the effect, the hypothesis that – in absence of heterogeneous nucleation sites – thickness or the nature of the support determines $N$, provides guidance on tuning the conditions of the 2D material surface to optimize the structure of the deposited material for a given application.

### 2.2. Effect of contamination, adsorbed water, and ripples on Au nucleation

So far, we have considered Au nucleation and diffusion processes on pristine suspended Gr. We now discuss the effects of extraneous nucleation sites, including how they alter Au morphology as well as nucleation density. To obtain the largest Au diffusivity and the intrinsic value of $N$, extraneous sites should of course be eliminated. In our samples the typical sources of extraneous nucleation are polymer residue from the transfer process, adsorbed water from exposure to ambient air, and wrinkles in the suspended membrane, especially for monolayer graphene.

We find that nucleation on clean Gr typically results in facetted and epitaxially aligned islands with diameters ~100 nm,[18, 19] as shown in Fig. 3a. This is consistent with previous reports for Au on bulk graphite.[28, 29] The diffraction pattern in Fig. 3b shows that the (111) plane of the face centered cubic Au structure is parallel to the Gr(0001) plane and the Au[220] direction is parallel to the Gr[10-10] direction. In line with previous studies,[54] these facetted islands[55] closely resemble the thermodynamically expected morphology, where facet size depends on surface and interface as expected from the Winterbottom construction for Au on Gr. The compact nature of the islands also shows that Au adatoms easily diffuse around the island once they attach to it.[9]

We find that this Au island morphology is extremely sensitive to surface contamination.[28, 30] Islands affected by contamination are readily visible in TEM images, as in Fig. 3a, where Au islands that nucleated on contamination are typically very small with a diameter on the order of 10 nm. Larger flat triangular nanocrystals surround these regions. We attribute the smaller islands to Au that landed within the contaminated region, with the larger islands supplied by diffusion of Au adatoms across the clean Gr. Fig. 3c shows that Au islands nucleated on and around contamination orient randomly with respect to the Gr support, unlike the epitaxial triangular islands. Hence, it is possible to estimate which Au islands nucleated on polymer residue, helping to optimize cleaning and deposition protocols.

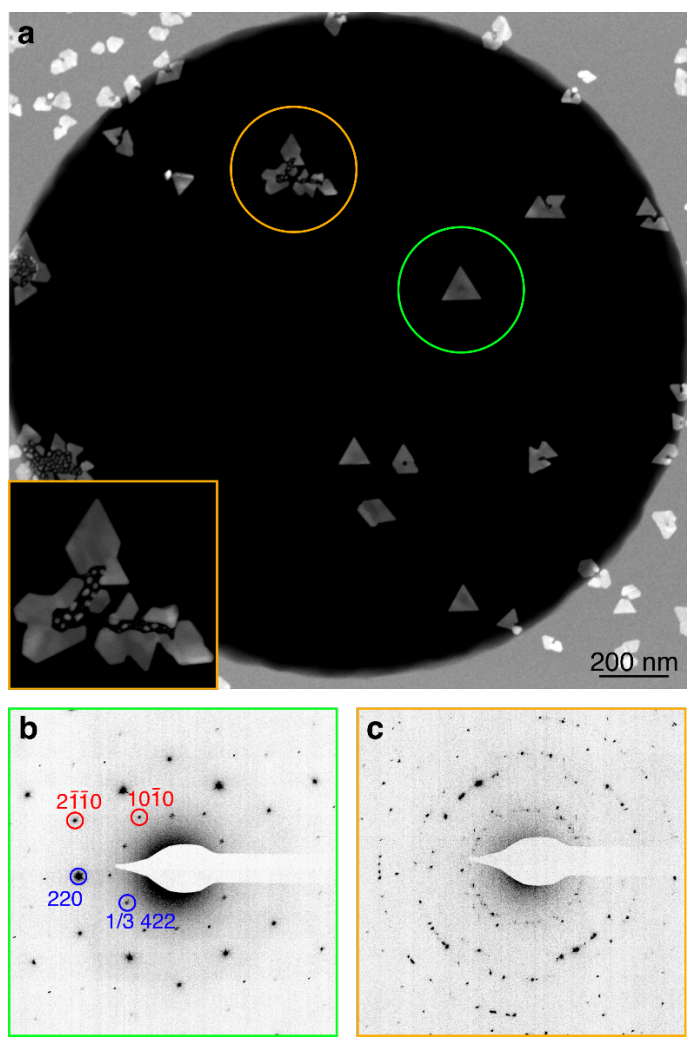

*Figure 3. Epitaxy of Au on clean suspended Gr. (a) TEM image of 3L Gr with deposited Au. The green and orange circle marks the extent of the selected area aperture used to capture the diffraction patterns in (b) and (c), respectively. The inset in (a) shows an enlarged view of the region indicated in orange. In (b) the red indices are for Gr and the blue are for Au.*

The samples we have discussed so far were annealed in UHV to temperatures above 500 °C immediately prior to deposition. Such vacuum annealing is presumed to remove polymer residues from the transfer process as well as adsorbed water. To examine annealing effects, in Fig. 4 we show the result of exfoliating Gr crystals onto an $SiO_2$ substrate then depositing Au without annealing. Au deposition on supported monolayer Gr (Fig. 4a) results in $N$=500 $\mu m^{-2}$, and the island morphology is less facetted as compared to our annealed samples. $N$ has been measured by others at ~1200 $\mu m^{-2}$ for room temperature deposition on $SiO_2$-supported monolayer Gr.[13] Our depositions take at place at around 100 °C and at lower rate, so our lower $N$ is consistent with this result. To evaluate how annealing affects deposition onto thicker layers, we show in Fig. 4b Au deposited on 128 nm thick graphite which was subsequently transferred to a TEM grid for imaging. The Au has the same epitaxial relation to the graphite as is shown in Fig. 3b, albeit with larger rotational variance, as evidenced by the diffuse diffraction spots shown in the inset of Fig. 4b. The value of $N$=150 $\mu m^{-2}$. Thus, both monolayer and bulk have significantly higher $N$ than the suspended samples (which were annealed), and the high $N$ cannot be attributed to the substrate because it is seen even on the thick 2D crystal.

These results show that annealing is a necessary step in the preparation of low-$N$ surfaces for epitaxy. Furthermore, we attribute the increased $N$ on the samples shown in Fig. 4 compared to their vacuum annealed counterparts in Fig. 1-3 to adsorbed water from ambient air which does not evaporate simply upon inserting the sample in UHV. This conclusion is in line with previous studies of Au nucleation on graphite.[30] It is interesting to note that standard metal evaporators do not allow for annealing the sample prior to metal deposition. However, given the striking effect on metal diffusion lengths and epitaxy, it would be interesting to study the effect of removing adsorbed water on the key physical properties of the metal-2D material interface, particularly the electrical contact resistance which is performance limiting for several applications.[56]

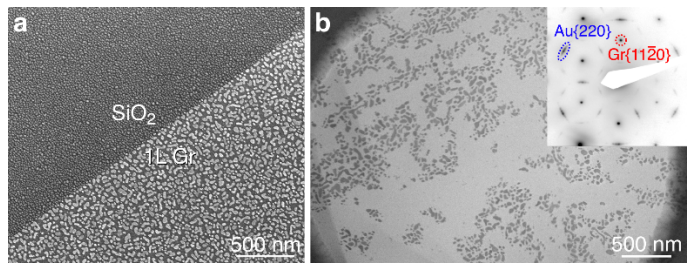

*Figure 4. Au deposition on un-annealed samples. (a) SEM image of Au evaporation on an as-exfoliated monolayer Gr crystal on $SiO_2$. (b) TEM image of Au deposited on 128 nm thick graphite while it was supported on $SiO_2$ and subsequently transferred to a TEM grid for TEM imaging. The inset shows a diffraction pattern with Gr indexes in red and Au in blue font.*

We finally discuss the origin of correlated heterogeneous nucleation sites. We occasionally have observed Au islands that have nucleated in lines as seen in Fig. 5. We excluded holes with such lines from our nucleation density analysis, as described in Methods. We see that several of these lines of Au are continuous across different holes although not directly visible on the supported regions due to the large nucleation density there. This suggests that these are static ripples, or possibly ripplocations in the Gr,[57, 58] arising due to the mechanical transfer process or thermal expansion during annealing. These lines were primarily observed on monolayer Gr, suggesting that the increased stiffness and stability of thicker crystals may hinder the formation of ripples. Control of nucleation sites therefore also requires attention to possible stresses caused by the transfer process. While grain boundaries could be the source of these lines, we note, however, that exfoliated graphene flakes are typically single crystalline. Our samples are single-crystalline, as seen from diffraction patterns measure over all suspended areas: Fig. S4 shows this approach for two samples. The low variation in rotation angle for the diffraction patterns (>0.5°) shows the single-crystalline nature of the graphene.[59, 60]

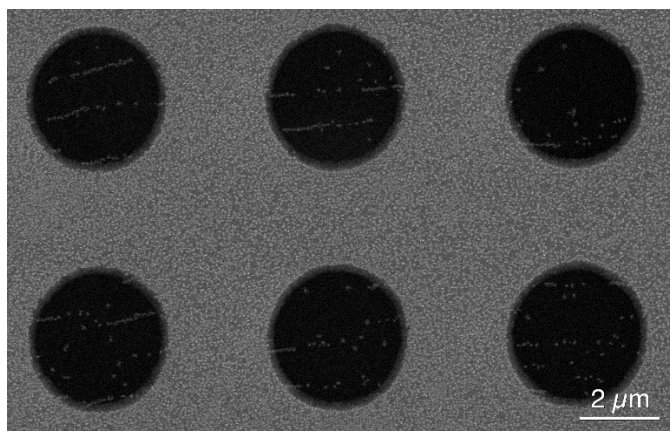

*Figure 5. Ripples in monolayer Gr. Scanning electron microscopy overview image of monolayer Gr on which 0.15 nm Au has been deposited.*

3. **Conclusions and Outlook**

By measuring Au nucleation and diffusion characteristics on Gr that has been prepared to minimize substrate effects and heterogeneous nucleation sites, we have shown that the nucleation density of Au on suspended Gr is significantly lower than on supported Gr. We furthermore find that the Au nucleation density on suspended Gr decreases with increasing layer thickness. The suppression of nucleation can be dramatic: for Gr crystals thicker than 5L it is possible to create suspended areas with zero Au nucleation events surrounded by dense arrays of islands in the supported regions. This indicates diffusion distances exceeding the hole radius of 1 $\mu$m, and an enhancement in diffusion constant of several orders of magnitude compared to the supported surface outside the hole.

These results are consistent with surface roughness being the main factor determining nucleation density and hence Au diffusivity in samples where extraneous nucleation sites are eliminated. This offers simple guidelines for tuning and achieving low nucleation densities: suspended samples are in general more flat compared to supported samples, and thicker samples are in general more flat than thinner samples. Sample thickness can also be increased by using hBN crystals beneath Gr as they act to reduce surface roughness in both suspended and supported samples. Our work also highlights the importance of cleaning and annealing the sample prior to Au deposition and UHV conditions during metal evaporation to minimize extrinsic nucleation sites such as hydrocarbons or adsorbed water; several methods may be useful in practice to provide the required surface condition.

Understanding the role of roughness, strain, contamination, and deposition conditions, particularly temperature, is necessary for full control of metal island and film morphology on Gr and provides opportunities for developing alternative approaches for device fabrication. We find that the difference in Au nucleation density for suspended and supported Gr can range from >1 $\mu m^{-2}$ for free-standing thicker crystals to ~500 $\mu m^{-2}$ for monolayer Gr on $SiO_2$. The controllability of nucleation density via the intrinsic diffusion properties of Au on graphene could be relevant for creating films with laterally varying properties by choosing the size of the suspended area and the layer number of the Gr. Our results further suggest the possibility of creating a coalesced film on supported Gr simultaneously with a discontinuous film on suspended Gr, without the need for lithographic patterning methods on the graphene surface. This may be the case if the lateral width of the film region is smaller than the intrinsic nucleation separation. The ultralong diffusion distance of suspended few-layer Gr also offers additional opportunities. A deliberate introduction of nucleation sites through means such as focused ion beam irradiation at a period smaller than the intrinsic nucleation separation could enable self-assembly of epitaxial islands only at specified locations and without the need for optical or electron beam lithography.[5] We anticipate that the diffusion and nucleation characteristics can further be controlled by changing the substrate temperature during evaporation; by tuning the conditions we would expect longer diffusion distances at elevated temperatures.

**Methods**

**Sample fabrication:** Gr was obtained by exfoliation of natural graphite (NaturGrafit GmbH, Germany) onto oxygen plasma-treated substrates of 90 nm thick $SiO_2$ on Si using 3M Magic scotch tape. We identified suitable crystals using optical microscopy, mapping regions of different thickness (layer number) from their contrast, see Fig. S1a. We then transferred the crystals onto either homemade TEM-compatible sample grids as shown in Fig. S1b or "location tagged" TEM grids from Norcada, Inc., Canada. The homemade grids consist of 300 nm thick SiN membrane windows with multiple ~3 $\mu$m diameter holes etched in the window. The Norcada TEM grids have a 200 nm thick SiN membrane with several 2 $\mu$m holes. Placement of the crystals on the grids was achieved using wedging transfer,[61] a method based on the use of cellulose acetate butyrate (CAB) as polymer handle. After transfer we bake the samples for 1 minute at 140 °C to improve the adhesion between the transferred crystals and the substrate. Then we dissolve the CAB in acetone for 15 minutes and dry the samples using critical point drying. This procedure has previously been shown to yield close to completely clean suspended Gr after vacuum annealing to temperatures >500 °C for >1 hour[62] and facetted Au nanoislands.[18,19]

**Au deposition:** Prior to evaporation we anneal the samples at 500 °C overnight in a vacuum chamber that also contains the Au evaporator. Without breaking vacuum, we evaporate Au using home-built thermal evaporators. Evaporation is assumed to occur only on one surface of the sample due to the ballistic flux and the long surface diffusion required to reach the back side of the sample. We measure the deposition rate using a quartz crystal microbalance and deposit typically 0.15 nm at a rate of 0.05 nm/min. However, the power needed to get this deposition rate varies 5-10 % between many experiments. Due to the small distance of about 1 cm between the sample and the evaporator we expect radiation from the evaporator to heat up the sample to some extent during evaporation: measurements using a thermocouple suggest a sample temperature of about 100 °C during evaporation. Hence, we expect some variation in the sample temperature across experiments which may explain the observed variation in $N$ seen in Fig. 2c. The pressure during evaporation is typically in the range $5\text{-}9 \times 10^{-9}$ Torr.

**Electron microscopy:** After deposition, we removed the samples from the vacuum system and imaged *ex situ* in either TEM, STEM, or SEM to quantify the deposition morphology at suspended regions of different layer thickness; supported regions could also be imaged through the silicon nitride at lower resolution using TEM or STEM. We used a JEOL 2010F TEM operated at 200 kV for TEM imaging and a FEI Themis STEM also operated at 200 kV for STEM imaging. The high kV used allowed for imaging through the silicon nitride; any damage caused to the 2D layer did not affect the nucleation density measurements. SEM imaging was performed in a Zeiss Gemini 450 SEM.

**Data analysis:** To calculate $N$ for each sample, we measure $N$ over individual suspended regions (holes in the membrane), then we compute the average $N$ for a sample as a function of Gr thickness. This simple count is modified by several factors: (1) A small number of Au islands in our samples appear to have nucleated on contamination, as shown in Fig. 3. The density of such islands does not appear to depend strongly on layer number. To avoid bias when calculating $N$ (Fig. 2), we count the cluster of Au islands nucleated on contamination as one island. We expect that the islands nucleated on contamination alter the overall value of $N$ slightly but not the layer thickness-dependent trend observed. (2) In cases where we observe Au nucleated along lines that may be step edges (see the bulk Gr in Fig. S2) or wrinkles in thin Gr (see Fig. 5), we exclude the hole from the count. (3) In one sample (Figure S2) the islands had dendritic shapes and the areas could not be calculated readily by our image processing algorithm. This is the sample that was excluded from Fig. 2d. Such variations in island geometry may be due to changes in experimental conditions due for example to small changes in evaporator power; we expect, in general, a higher evaporation rate or lower substrate temperature to result in dendritic growth.[42]

On Gr supported by SiN, the layer dependence of nucleation density is obscured by geometric factors as described in SI Section 2.

In considering factors that may affect the data analysis, particularly the comparison of nucleation density in suspended and supported regions, we assume uniform flux of Au over the sample area. We measured data at or near holes that were fully covered with Gr, where the Gr extended several micrometers beyond the holes. We further did not see effects of proximity to the Gr edge. We do not expect sticking probabilities of Au to be different in suspended and supported regions since thermal evaporation sticking probabilities are close to 1 for most surfaces. We also do not expect suspended regions to be at a lower temperature due to the low heat capacity and high thermal conductivity of Gr.

ACKNOWLEDGMENT

J.D.T. acknowledges support from Independent Research Fund Denmark though Grant Number 9035-00006B. K.R. acknowledges funding and support from a MIT MathWorks Engineering Fellowship and ExxonMobil Research and Engineering Company through the MIT Energy Initiative. J.K. acknowledges support by the Alexander von Humboldt foundation. The authors acknowledge use of the MRSEC Shared Experimental Facilities at MIT, supported by the National Science Foundation under award number DMR-1419807, and use of the MIT.nano Characterization Facilities. The authors would like to acknowledge Michael Tarkanian for help in manufacturing TEM sample holders, Prof. Ju Li for discussions on ripplocations, and Prof. Jeehwan Kim for equipment access.

# Supplementary information:

# Suspended graphene membranes to control Au nucleation and growth

Joachim Dahl Thomsen[1], Kate Reidy[1], Thang Pham[1], Julian Klein[1], Anna Osherov[2], Rami Dana[1], Frances M. Ross[1]

[1]Department of Materials Science and Engineering, Massachusetts Institute of Technology, Cambridge, Massachusetts 02139, USA

[2]Department of Electrical Engineering and Computer Science, Massachusetts Institute of Technology, Cambridge, Massachusetts 02139, USA

## Section 1. Sample overview

Figure S1a shows an optical microscopy image of as-exfoliated graphene (Gr) with different layer thicknesses across the crystal, as identified by the contrast to the bare SiO$_2$ substrate (see inset). Figure S1b shows the crystal after transfer onto a homemade TEM-compatible sample grids.

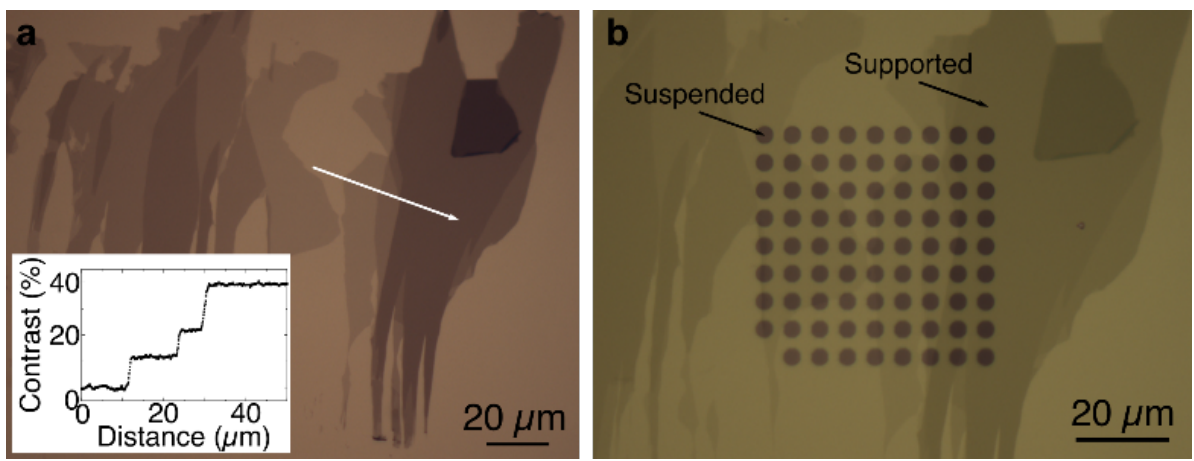

*Figure S1:* Sample overview. (a) Optical microscopy of an as-exfoliated Gr crystal with different thicknesses. The inset shows the contrast, $C(x) = (I_{BG} - I(x))/I_{BG}$, where $I_{BG}$ is the average grey value of the bare SiO$_2$ substrate, and $I(x)$ is the grey value along the white arrow in (a). The contrast increases in multiples of about 11%, showing that the sample consists of 1, 2, and 4L thick Gr. (b) The crystal from (a) transferred to our homemade TEM sample grids. The original optical image from (a) is overlaid on the image of the transferred crystal and made transparent. This makes it easier to identify the layer thickness over each hole.

## Section 2. Nucleation density on SiN-supported Gr

Figure S2 and S3 show overview images of two samples. Figure S2a-c and Fig. S3a,b show the Gr before and after transfer, while Fig. S2d and Fig. S3c are stitched TEM and SEM images, respectively, showing the entire sample. In Fig. S2d Au nanocrystals appear as bright speckles both on the suspended and supported Gr, while in Fig. S3c the Au nanocrystals appear as dark speckles due to the bright field imaging mode used. The nucleation density in the suspended regions varies consistently as a function of layer thickness, see Fig. 2c in the main text. However, we find that that Au density on the SiN-supported Gr is inhomogeneous across the sample. For instance, in Fig. S3c there are two regions with 4L Gr. The lower of the two regions has a nearly coalesced film of Au while the nucleation density in the upper region is sparser. Also, within the 8L region in Fig. S2d, the upper part has a higher nucleation density than the lower part. This makes the determination of nucleation density layer dependency challenging for supported Gr. We believe that geometrical effects may have a particularly strong influence on Au diffusion on supported Gr in our samples. The placement of the Gr crystal across the holes, and the positional relation between the step edges, which act as a sink and possible blocking site for Au by nucleating a row of Au islands, may cause longer range spatial variations in nucleation density on supported Gr that drown out the layer dependence.

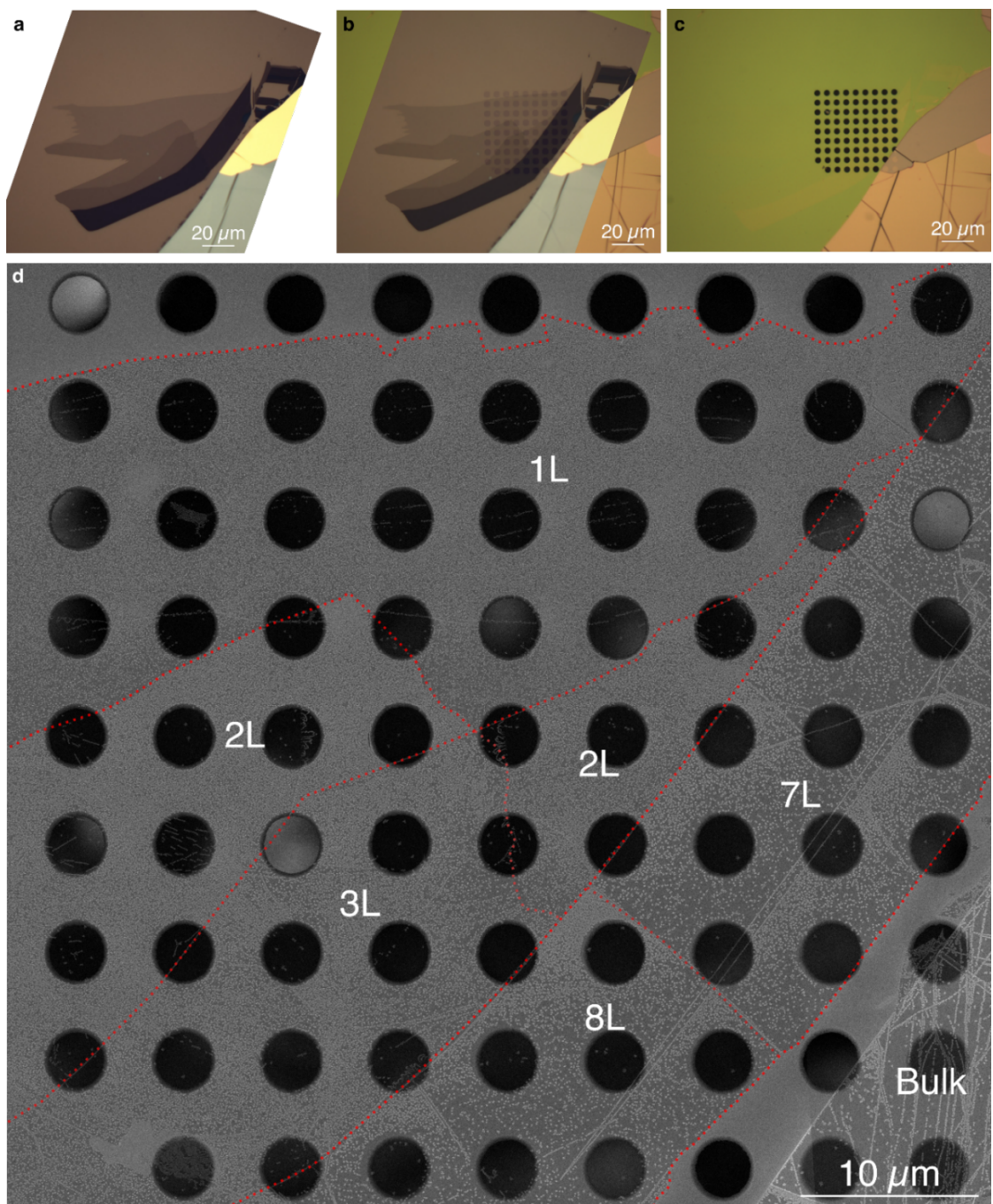

***Figure S2***: Overview of the sample shown in Fig. 2c with square, dark blue data points. (a) Optical image of the Gr crystal before transfer. (b) The optical image overlaid the sample shown in (c) and made transparent. (c) Optical image of the sample after transfer of the Gr. (d) Montage of SEM images of the sample. The square dark blue data points in Fig. 2c are from the sample shown in Fig. S2. In this sample, the Au islands were very irregular and dendritically shaped which prevented our image processing algorithm from correctly measuring island areas.

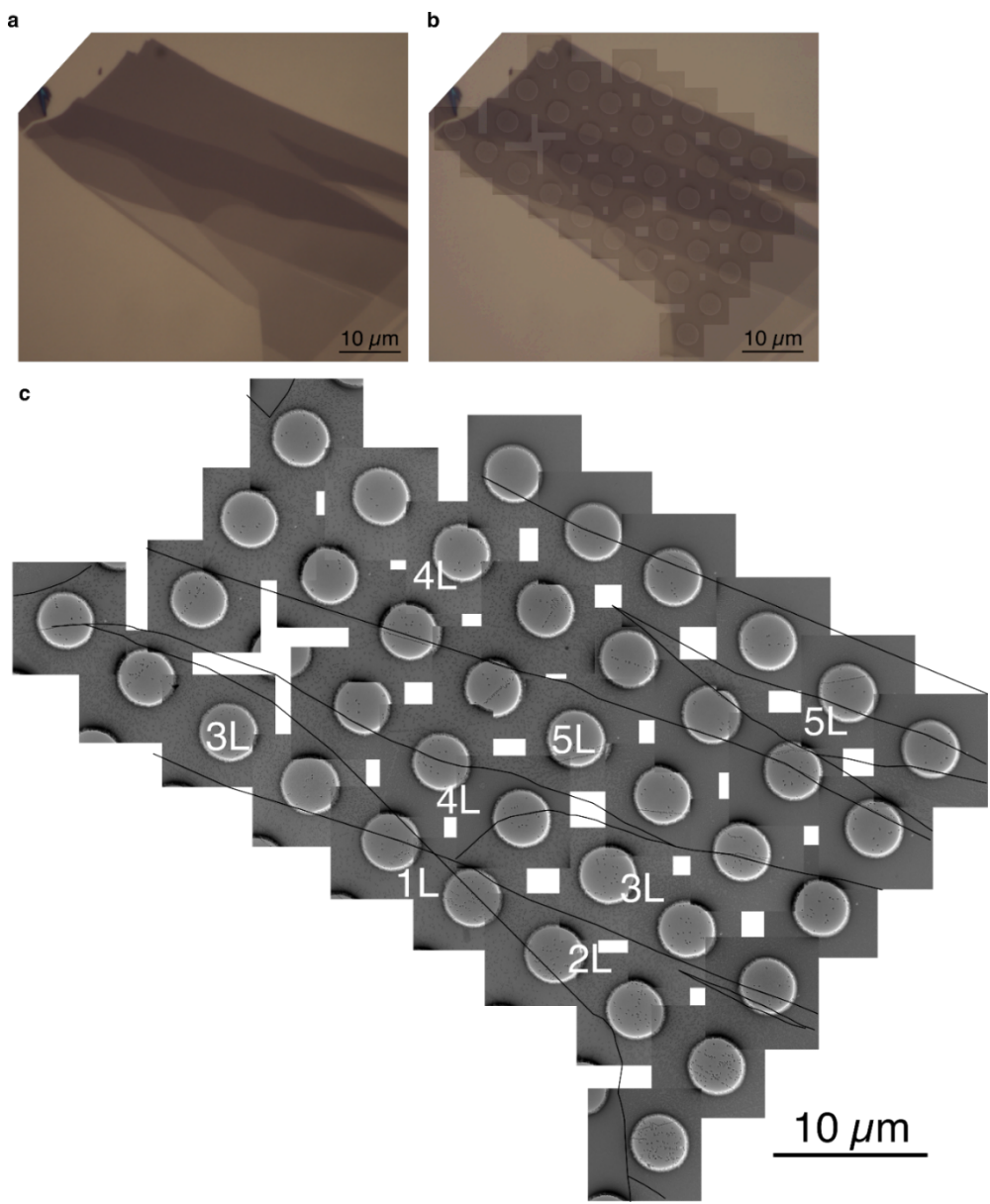

***Figure S3***: *Overview of the sample shown in Fig. 2c shown with round, blue data points. (a) Optical image of the Gr crystal before transfer. (b) The optical image overlaid on the image shown in (c) and made transparent. (c) Montage of bright field TEM images of the sample.*

## Section 3. Estimation of diffusion constants and diffusion barriers

In line with the methods described in the main text, Supplementary Table 1 summarizes the estimates in the changes of diffusion constant and diffusion barrier. All values are compared to suspended graphite.

***Supplementary Table 1:*** *Estimates for changes in diffusion constant and diffusion barrier for different types of graphene samples. Values for diffusion constant are fractions of $D_\infty$, while values for diffusion barrier is the increase in eV compared to $E_{d,\infty}$. The increase in $E_d$ is calculated from the equation $\Delta E_d = \left(\frac{i+2.5}{i}\right) \cdot kT \cdot \ln\left(\frac{N}{N_\infty}\right)$, where N is the nucleation density of Au on type of graphene sample given in the table and $N_\infty=0.8$ $\mu m^{-2}$ is the nucleation on graphite.*

| Critical nucleus size assumed, i | 1L, suspended | | 3L, suspended | | 1L, supported on SiO₂, unannealed | | 3L, supported on SiN, annealed | | Graphite, supported on SiO₂, unannealed | |
|---|---|---|---|---|---|---|---|---|---|---|
| | D | $E_d$ | D | $E_d$ | D | $E_d$ | D | $E_d$ | D | $E_d$ |
| 1 | 3·10⁻⁴ | 0.21 | 9·10⁻² | 0.06 | 2·10⁻¹⁰ | 0.59 | 3·10⁻⁸ | 0.45 | 1·10⁻⁸ | 0.48 |
| 2 | 6·10⁻³ | 0.13 | 2·10⁻¹ | 0.04 | 5·10⁻⁷ | 0.38 | 2·10⁻⁵ | 0.29 | 8·10⁻⁶ | 0.31 |

## Section 4. Measurements of sample mono-crystallinity

To confirm the single crystalline nature of our samples we measured diffraction patterns over several suspended regions for two of the samples. Results are shown in Fig. S4.

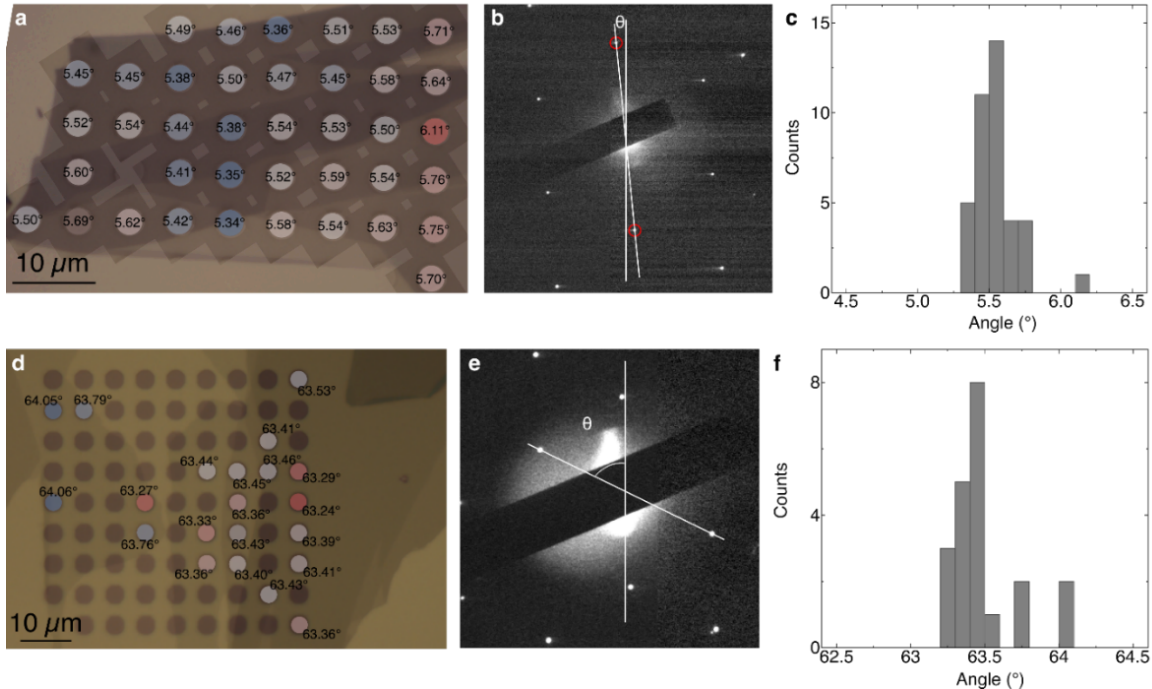

***Figure S4****: Measurements of diffraction patterns over several suspended regions in two samples. (a, d) optical microscope images overlaid with stitched TEM images of the sample. The image in (a) is the same image as shown in Fig. S3b. The suspended regions are shaded in a color where white indicates that the diffraction pattern is rotated close to the average, and blue and red*

*regions have rotations that are the rotated furthest away from the average. (b, e) Diffraction patterns from one of the suspended regions in each sample, showing how the rotation angle was measured. (c, f) Histograms of the rotation angles for each sample.*